\documentclass[journal, onecolumn]{IEEEtran}

\usepackage[T1]{fontenc}
\usepackage{graphicx}
\usepackage{geometry}
\usepackage{authblk}
\usepackage{soul}
\usepackage{amsmath}
\usepackage{lineno}
\usepackage{setspace}
\usepackage{xcolor}

\usepackage{amsmath}
\usepackage{alphabeta}

\usepackage{cite}

\newcommand{\jpet}{\mbox{J-PET}}
\newcommand{\opsanh}{$\text{o-Ps}{\to}3\gamma$}
\newcommand{\upbeta}{\text{\beta}}
\let\gamma\relax
\DeclareMathOperator{\gamma}{\text{\gamma}} 

\title{Testing CPT symmetry in ortho-positronium decays
with positronium annihilation tomography}

\author[1,2,*]{P.~Moskal}
\author[1,2,**]{A.~Gajos}
\author[1]{M.~Mohammed}
\author[1,2]{J.~Chhokar}
\author[1,2]{N.~Chug}
\author[3]{C.~Curceanu}
\author[1,2]{E.~Czerwi{\'n}ski}
\author[1,2]{M.~Dadgar}
\author[1,2]{K.~Dulski}
\author[4]{M.~Gorgol}
\author[5]{J.~Goworek}
\author[6]{B.~C.~Hiesmayr}
\author[4]{B.~Jasi{\'n}ska}
\author[1]{K.~Kacprzak}
\author[1,2]{\L{}.~Kap\l{}on}
\author[1,2]{H.~Karimi}
\author[1]{D.~Kisielewska}
\author[7]{K.~Klimaszewski}
\author[1,2]{G.~Korcyl}
\author[7]{P.~Kowalski}
\author[1,2]{N.~Krawczyk}
\author[8]{W.~Krzemie{\'n}}
\author[1]{T.~Kozik}
\author[1,2]{E.~Kubicz}
\author[1,2]{S.~Nied{\'z}wiecki}
\author[1,2]{S. Parzych}
\author[1,2]{M.~Pawlik-Nied{\'z}wiecka}
\author[7]{L.~Raczy{\'n}ski}
\author[1,2]{J.~Raj}
\author[1,2]{S.~Sharma}
\author[1,2]{Shivani}
\author[7]{R.Y.~Shopa}
\author[5]{A.~Sienkiewicz}
\author[1,2]{M.~Silarski}
\author[1,3]{M.~Skurzok}
\author[1,2]{E.~{\L{}}.~St\k{e}pie{\'n}}
\author[1,2]{F.~Tayefi}
\author[7]{W.~Wi{\'s}licki}
\affil[1]{Faculty of Physics, Astronomy and Applied Computer Science, Jagiellonian University,  S.~Łojasiewicza 11, 30-348 Kraków, Poland}
\affil[2]{Total-Body Jagiellonian-PET Laboratory, Jagiellonian University, Poland}
\affil[3]{INFN, Laboratori Nazionali di Frascati CP 13,  Via E. Fermi 40, 00044, Frascati, Italy}
\affil[4]{Department of Nuclear Methods, Institute of Physics, Maria Curie-Sk{\l}odowska University, Pl.~M.~Curie-Sk{\l}odowskiej~1, 20-031 Lublin, Poland}
\affil[5]{Faculty of Chemistry, Institute of Chemical Sciences, Maria Curie-Sk{\l}odowska University, Pl.~M.~Curie-Sk{\l}odowskiej~3, 20-031 Lublin, Poland}
\affil[6]{Faculty of Physics, University of Vienna  Boltzmanngasse 5, 1090 Vienna, Austria}
\affil[7]{Department of Complex Systems, National Centre for Nuclear Research,  05-400 Otwock-Świerk, Poland}
\affil[8]{High Energy Department, National Centre for Nuclear Research,  05-400 Otwock-Świerk, Poland}
\affil[*]{corresponding author, e-mail: p.moskal@uj.edu.pl}
\affil[**]{corresponding author, e-mail: aleksander.gajos@uj.edu.pl}

\begin{document} 
\maketitle
\onehalfspacing

\begin{abstract}
  Charged lepton system symmetry under combined charge, parity, and time-reversal transformation (CPT) remains scarcely tested.
  Despite stringent quantum-electrodynamic limits, discrepancies in predictions for the electron--positron bound state (positronium atom)
  motivate further investigation, including fundamental symmetry tests.
  While CPT noninvariance effects could be manifested
  in non-vanishing angular correlations between final-state photons and spin of annihilating positronium,
  measurements were previously limited by knowledge of the latter.
  Here, we demonstrate tomographic reconstruction techniques applied
  to three-photon annihilations of ortho-positronium atoms to estimate
  their spin polarisation without magnetic field or polarised positronium source.
  {
    We use a plastic-scintillator-based positron-emission-tomography scanner
    to record ortho-positronium~(o-Ps) annihilations
    with single-event estimation of o-Ps spin
    and determine the complete spectrum of an angular correlation operator
    sensitive to CPT-violating effects.
  }
  We find no violation at the precision level of $10^{-4}$, with an over threefold improvement on the previous measurement.  
  \end{abstract}


\section*{{Introduction}}\label{sec:introduction}
The symmetry under the combined operations of charge conjugation~(C), spatial inversion~(P), and reversal in time~(T), referred to as CPT invariance, is the last of the fundamental discrete symmetries, a violation of which has never been observed.
While numerous experimental tests have been carried out using baryonic matter systems (such as, neutral mesons that provide stringent limits~\cite{RevModPhys.83.11,Mavromatos:2005bg,Babusci:2013gda,RevModPhys.87.165}), and
significant efforts have recently been made towards comparing the properties of hydrogen and anti-hydrogen atoms~\cite{Kuroda:2014,Fitzakerley:2016mbn,Ahmadi:2017gwe,Ahmadi:2016fir,Ahmadi:2018eca,Ahmadi:2018dqm},
certain CPT symmetry tests are also conceivable with purely leptonic systems.

Studies of neutrino oscillations in long-baseline experiments are prominent candidates in this sector. Discrepancies in mass and mixing parameters between neutrinos and antineutrinos would indicate CPT violation~\cite{Barenboim:2017ewj, PhysRevD.96.011102},
and while it has been argued that the sensitivity of such experiments could surpass even that of the neutral kaon system~\cite{Barenboim:2017ewj}, recent results agree with CPT conservation~\cite{abe2020t2k}.

Systems consisting of charged leptons, in contrast, are governed by quantum electrodynamics (QED), supported by experimental evidence of unparalleled precision.
The most recent measurements of the electron anomalous magnetic moment~\cite{PhysRevLett.100.120801},
and its QED predictions based on the recent precise
measurement of the fine structure constant~\cite{Parker:2018vye}
agree to 1 part in $10^{12}$,
indicating that no new CPT-violating interactions coupling to the electron could be observed to this level of precision~\cite{Bass:2019ibo}.

Nonetheless, the recently discovered deviations from QED predictions 
in the fine structure of the positronium atom, that is, the bound state of an electron and positron~\cite{PhysRevLett.125.073002}, 
suggest that diverse experimental studies of charged lepton systems should not be hastily abandoned.
In fact, as the lightest matter--antimatter bound states,
positronium atoms are considered in the context of CPT tests through
searches for Lorentz noninvariance.
Such tests, based on atomic spectroscopy of positronium have been recently proposed in the theoretical framework of Standard-Model Extension~\cite{Kostelecky:2015nma, Vargas:2019swg}.

A complementary, model-independent approach to searching for new CPT-violating interactions
in the positronium system is constituted by searching for CPT-prohibited angular correlations in the annihilation of ortho-positronium atoms into three photons (\opsanh{}).
This approach, first reported in 1988~\cite{Arbic:1988pv},
has seen relatively little activity in recent years,
with the best result to date achieving a sensitivity of $3\times 10^{-3}$~\cite{cpt_positronium}.  

In this work, we {present an experimental approach}
to search for a CPT-violating angular correlation
in the three-photon annihilation of ortho-positronium atoms.
We employ a positron-emission-tomography (PET) imaging device
and {demonstrate that} three-photon annihilations of positronium
can be reconstructed in a large volume,
which {allows for estimating} its spin polarisation without
the use of an external magnetic field.
Using this technique, we measure the expectation value
of a CPT-prohibited angular correlation $\mathbf{S}\cdot(\mathbf{k}_1\times\mathbf{k}_2)$, improving the sensitivity of the previous result~\cite{cpt_positronium} by over a factor of three.
Notably, {the \opsanh{} reconstruction} technique presented here
additionally opens the possibility of imaging
of excited long-lived positronium states in studies of their gravitational fall as a matter--antimatter system~\cite{Mariazzi:2020bgc}.

\section*{{Results}}\label{sec:results}
\subsection*{CPT-prohibited angular correlations in ortho-positronium annihilations}\label{sec:corr-def}

The decaying ortho-positronium state is described by its unit spin
 along a given axis $\mathbf{S}$,
whereas the final state of this annihilation is characterised by the momenta of the three photons $\mathbf{k}_i, \; i=1,2,3$, where we assume an ordering of $|\mathbf{k}_1| > |\mathbf{k}_2| > |\mathbf{k}_3|$.
Certain angular correlation operators constructed using these observables are sensitive to the effects of violation of T, CP, or combined CPT symmetry if the corresponding operators are odd under the given symmetry transformation~\cite{Bernreuther:1988tt, Adkins:2010pd}.
Notably, these symmetry-violating correlations are not limited to positronium, and a similar approach has been employed to search for T noninvariance in the decays of Z\textsuperscript{0} bosons into hadronic jets~\cite{PhysRevLett.75.4173}.

Here, we focus on a study of the following operator:
\begin{equation}
  \label{eq:operator}
  O_{\rm CPT} = \mathbf{S}\cdot(\mathbf{k}_1\times\mathbf{k}_2) / |\mathbf{k}_1\times\mathbf{k}_2 | = {\rm cos}\theta,
\end{equation}
expressing the angular correlation $\theta$ between the ortho-positronium spin and the orientation of the decay plane spanned by the three annihilation photons.
This correlation operator constitutes a robust CPT-violation-sensitive observable
defined through the fundamental geometry of an annihilation event,
independent of particular kinematical configurations.
The definition of the annihilation plane orientation through the two most energetic photons is merely a convenient convention and thus does not impose a bias on the space of kinematical configurations of three-photon annihilations,
a problem affecting other conceivable angular correlations for o-Ps$\to 3\gamma$ events.
Consequently, the choice of this observable for the CPT test minimises spurious asymmetries originating from the detector geometry that could mimic a violation~\cite{Gajos:2020symmetry}.
Moreover, due to the neutrality of the final state
of the \opsanh{} process, radiative corrections that could cause fake CPT asymmetry,
thus posing a natural limit on the sensitivity of such a test, 
are only expected at a precision level of $10^{-9}$~\cite{Bernreuther:1988tt, Arbic:1988pv},
thus leaving significant exploratory potential.

\subsection*{Previous measurements}\label{sec:prev-meas}

Measurement of angular correlations in \opsanh{} decays requires knowledge of the positronium spin as well as the recording of the momenta of the annihilation photons. Previous studies of CP- and CPT-violating operators achieved the former by either utilising a polarised positron beam~\cite{Arbic:1988pv} or polarising an o-Ps sample in an external magnetic field.
The latter approach has been commonly used in studies of the CP-violation-sensitive correlation $(\mathbf{S}\cdot \mathbf{k}_1)(\mathbf{S}\cdot\mathbf{k}_1\times\mathbf{k}_2)$, 
which has a sensitivity to symmetry violation that demands specific tensor polarisation of ortho-positronium~\cite{Skalsey:1991vt, cp_positronium}.
Notably, these experiments simultaneously recorded photons in only a single annihilation plane so that the magnetic-field-generation setup did not overlap with the detection devices. In turn, the measurements were limited to a scalar asymmetry between two opposite configurations integrated over possible angles.
Moreover, the need to reconfigure the detection setup to detect opposite decay geometries accounted for a significant source of systematic uncertainty in these measurements~\cite{Skalsey:1991vt, cp_positronium}.

The most recent search for CPT violation using the $\cos\theta$ correlation, therefore, used the intrinsic polarisation of positrons from $\upbeta^+$ decay combined with limiting the direction of their emission, which allowed for the use of a high-angular-acceptance photon detector capable of recording multiple scattering planes simultaneously. 
Thus, the first search for asymmetry in the distribution of the $\cos\theta$ operator could be performed~\cite{cpt_positronium}.
The statistical polarisation of o-Ps was obtained by allowing positrons from a $^{22}$Ne or $^{68}$Ge source to thermalise and form positronium only in a hemisphere of aerogel with the source at its centre, elucidating the positron emission direction with $\Delta\vartheta=90^{\circ}$.
The effective polarisation of the positrons was therefore given by
$\frac{\upsilon}{c} \frac{1}{2}(1+\cos(\Delta\vartheta))$,
where $\frac{\upsilon}{c}$ is the intrinsic polarisation of positrons from $\upbeta^{+}$ decay along the direction of emission, and the remainder is a geometrical factor limiting the directional e$^+$ polarisation, in this case to a cone with a $2\Delta\vartheta$ opening angle~\cite{Coleman}. 
Owing to the hemisphere used as the positronium production medium and the dimensions of the e$^+$ trigger detector used, this geometrical factor amounted to 0.686.
Moreover, despite the spherical symmetry of the detector, the chosen axis of the hemisphere required the source setup to be rotated in subsequent phases of the measurement to reduce systematic effects~\cite{cpt_positronium}.

\subsection*{Positronium spin estimation and reconstruction of \opsanh{} events in \jpet{}}\label{sec:reco}
Our study was performed using the Jagiellonian Positron Emission Tomograph (\jpet{}) detector, which was conceived as the first PET imaging device based on plastic scintillators~\cite{Moskal:2014sra, Niedzwiecki:2017nka}.
In addition to exploring the path towards a cost-effective total-body PET scanner~\cite{pet_clinics, Niedzwiecki:2017nka, total_body_pet} and imaging with properties of $e^{+}e^{-}$ bound states produced during scans of living organisms~\cite{Moskal:2018wfc,nature-review-physics2019,Moskal:2019nqk}
as well as positron annihilation lifetime spectroscopy~\cite{Dulski:2018huc},
\jpet{} constitutes a robust photon detector well suited to the study of angular correlations in \opsanh{} annihilations~\cite{moskal_potential, Dulski:2020pqi}.

\begin{figure}[h!]
  \centering
  \includegraphics[width=0.9\textwidth]{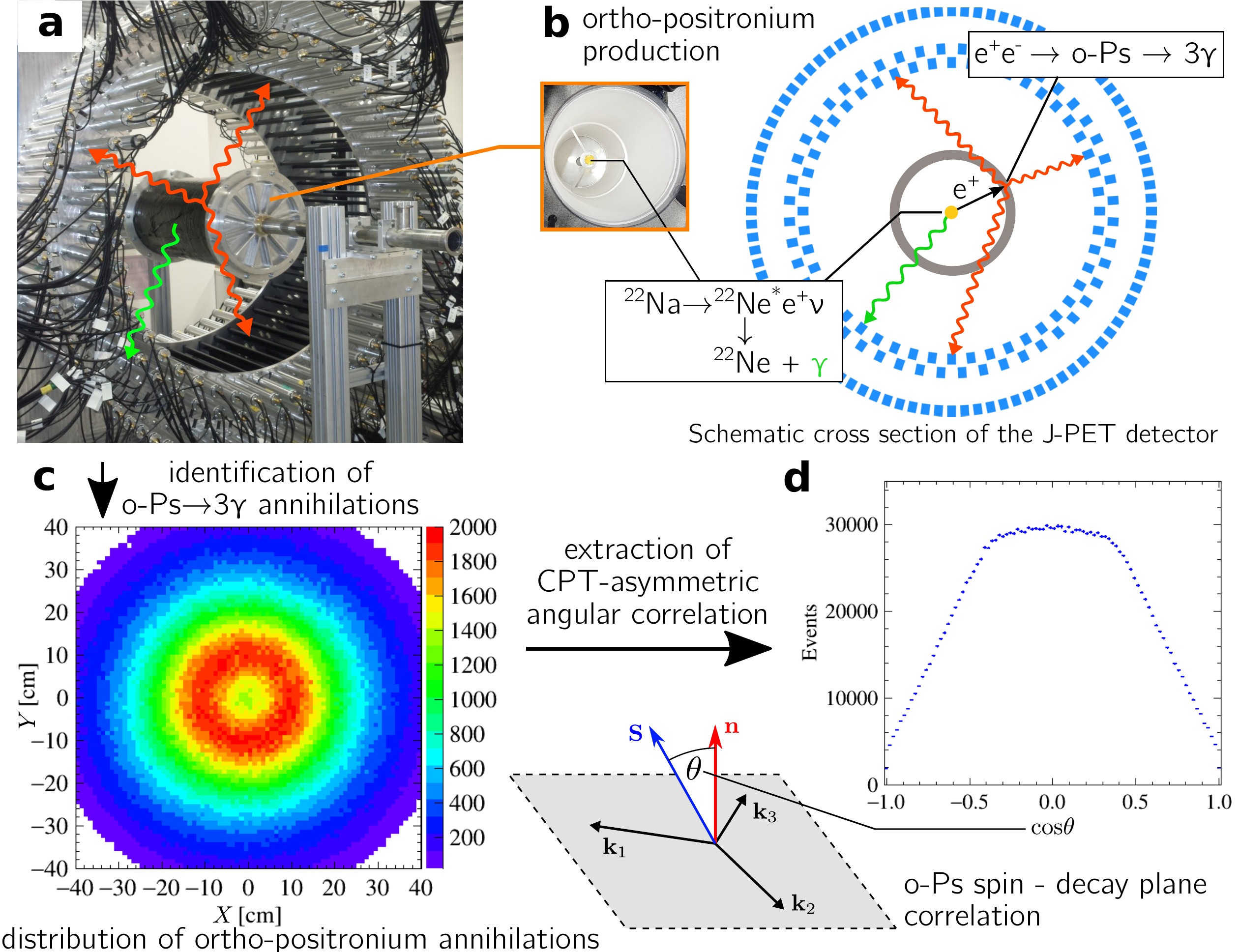}
  \caption{{\textbf{Methodology of the measurement of CPT-violation-sensitive angular correlation in three-photon annihilations of ortho-positronium.}}
    \textbf{a}~Photograph of the \jpet{} tomograph with the cylindrical positronium production chamber.
    Plastic scintillators~(black strips) record photons produced in \opsanh{} events~(red arrows) and a prompt photon from $\upbeta^+$ emitter de-excitation~(green arrow).
    \textbf{b}~Photograph of the interior of the positronium production chamber and a schematic view of the transverse cross section of the \jpet{} tomograph. Plastic scintillators (with rectangular cross sections~(blue)) form three coaxial rings.
    Annihilating o-Ps atoms are produced by positrons from a \textsuperscript{22}Na $\upbeta^+$~source in the centre of a cylindrical vacuum chamber~(grey), the walls of which are coated with porous silica, where positrons thermalise and form positronia.
    \textbf{c}~Reconstructed spatial locations of the identified \opsanh{} events
    {obtained using} three-photon annihilations, allowing us to reproduce a tomographic image of the vacuum chamber.
    The maximum density distribution is in a ring with a radius of 12~cm, equal to the radius of the positronium production chamber.
    \textbf{d}~For every selected event, the o-Ps spin axis is reconstructed, allowing calculation of the cosine of the angle between the spin~($\mathbf{S}$) and annihilation plane orientation~($\mathbf{n}=\mathbf{k}_1\times\mathbf{k}_2$), the distribution of which is sensitive to CPT-violating asymmetries.
    The histogram presents the determined $\textnormal{cos}\theta$ distribution for $1.9 \times 10^6$ identified \opsanh{} events.
  }
  \label{fig:panel}
\end{figure}

Figure~\ref{fig:panel} presents the methodology used in this work for studying \opsanh{} events with positronium spin control without the use of an external magnetic field. 
Annihilation photons are detected through their Compton interaction in strips of plastic scintillators with an interaction time resolution at the level of 250~ps and an angular resolution of approximately $1^{\circ}$~(Fig.~\ref{fig:panel}a).

Data acquisition operates in a triggerless mode~\cite{Korcyl:2016pmt,korcyl_ieee} to minimise the initial sample bias for discrete symmetry tests and searches for rare positronium decays~\cite{Gajos:2018wyi}.
A detailed description of the \jpet{} detector can be found {in the ``Methods'' section}.

To estimate the ortho-positronium spin by utilising the imaging capabilities of the \jpet{} tomograph,
we devised an extension of the spin control method
used in Ref.~\cite{cpt_positronium}
based on allowing positrons from a point-like $\upbeta^{+}$ source
to form o-Ps in a large volume following the symmetry of our photon detector. 
We used a $^{22}$Na source placed in the centre of a cylindrical vacuum chamber, the walls of which were coated with a layer of porous medium, enhancing positronium formation~(Fig.~\ref{fig:panel}b).

The positron source is prepared as a micro-droplet of liquid \textsuperscript{22}NaCl, evaporated and closed in a thin 7~$\textnormal{\mu}$m Kapton foil with density of $\sim$1.5~g~cm\textsuperscript{-3} resulting in areal density of $\sim$1~mg~cm\textsuperscript{-2}.  Therefore, the scattering and depolarization of positrons in the source material is negligible with respect to the 8~\% polarisation loss estimated for the 3~mm thick target material with the density of 0.32~g~cm\textsuperscript{-3} resulting in areal density of $\sim$100~mg~cm\textsuperscript{-2}.

The porous material layer, consisting of mesoporous silica and gypsum~(similar to that used in thin-layer chromatography), allowed for preparation of a highly porous self-supporting structure of extensive size.
The usage of gypsum enhances binding of the silica gel
and its expansion during setting allows for better packing of the material on a concave surface,
such as the inner wall of the cylindrical chamber.
The porous medium was prepared by multi-stage deposition through the evaporation of a silica--gypsum suspension in an alcohol solution distributed on the inner surface of the cylinder. 

The thickness of the obtained silica coating was approximately 3~mm, well above the maximum implantation depth expected for unmoderated positrons from~$^{22}$Na~\cite{Dryzek:2005}.
The vacuum chamber walls consisted of 4~mm thick polycarbonate, allowing transmission of photons
from positronium annihilation at the level of 90~\%~\cite{Gorgol:2020acta}.
A pressure of $10^{-3}$~Pa was maintained inside the chamber to reduce the scattering of positrons travelling from the source to the positronium production medium on the chamber walls~\cite{gajos_slopos}.

It is important to note that smearing of annihilation positions with respect to the porous material location caused by positronia escaping from the open pores is significantly smaller in comparison to the present resolution of the reconstructed annihilation points which amounts to about 8~cm.

To maintain the symmetry of the setup, we did not limit the e$^+$ emission direction to any solid angle besides the limits of the chamber. 
Instead, the positron flight direction was estimated separately for every recorded three-photon event using the position of the $3\gamma$ annihilation vertex reconstructed with {a trilateration-based technique}~\cite{gajos_gps, Gajos:2018wyi}.

Initial identification of \opsanh{} event candidates in the \jpet{} data was based on photon energy deposition
measured through time-over-thresholds (TOTs) of pairs of photomultiplier signals 
recorded at two ends of a scintillator strip for every registered photon interaction~\cite{Sharma:2019vrv,Sharma:2020}. 
As Compton interaction in the \jpet{} plastic scintillators yielded a continuous energy deposition spectrum, 
candidates for the identification of photons from an \opsanh{} annihilation were recognised using a TOT window located below the Compton edge of 511~keV photons, as indicated in Fig.~\ref{fig:meth:tot}.
The fast response of the plastic scintillators allowed a fine coincidence time window of 2.5~ns to be employed for the identification of $3\gamma$ event candidates.

The high angular resolution of the detector at the level of 1$^{\circ}$, coupled with exclusive registration of \opsanh{} decays, compensates for the lack of direct measurement of the photon energy.
Their momenta are completely reconstructed using the recorded points of Compton interactions in the detector.
Subsequently, the directions of $\mathbf{k}_1$, $\mathbf{k}_2$, and $\mathbf{k}_3$ span between the $\gamma$ recording
points and the annihilation point reconstructed using the latter
combined with the corresponding interaction times.
Consequently, the knowledge of the relative angles between the photon momenta can be used to unambiguously determine their energies~\cite{daria_epjc}.

Further selection of \opsanh{} events is based on the reconstructed photon energies and relative angles between their momentum vectors.
Moreover, the background originating from secondary Compton-scattered photons recorded in the detector
is discriminated by testing a hypothesis of a photon travelling directly between all pairs of detected $\gamma$~interactions~{(see ``Methods'' for details)}.

The identified sample of three-photon annihilations of ortho-positronium atoms constitutes
the first measurement of such events taking place in a medium of extensive size and featuring 
annihilation point reconstruction, which allows for determination of a three-photon tomographic image
of the annihilation chamber using the spatial density of \opsanh{} events~(Fig.~\ref{fig:panel}c).
In addition to the studies discussed in this work, this newly explored tomographic capability 
opens prospects for
{non-standard medical imaging modalities}
with spatially resolved determination of positronium properties
~\cite{Moskal:2018wfc,nature-review-physics2019,Moskal:2019nqk,pet_clinics}.

\subsection*{Search for CPT-violating angular correlations in \opsanh{}}\label{sec:cpt}

In August 2018, we collected a total of $7.3 \times 10^6$ event candidates
in a continuous 26-day measurement using the described setup with a 10~MBq \textsuperscript{22}Na positron source.
Figure~\ref{fig:panel}c presents the distribution of reconstructed \opsanh{} annihilation points in the transverse plane of the detector 
where the location of the porous silica allowing for positronium formation on the walls of the vacuum chamber is visible as the region with the highest density of reconstructed points.
Notably, this distribution constitutes the first image of an extensive-size object obtained using \opsanh{} annihilations.

To ensure proper reconstruction of the ortho-positronium spin, the recorded photon interaction times were varied within their uncertainties so as to minimise the discrepancy between $\rho=\sqrt{X^2+Y^2}$ of the reconstructed annihilation point and the radius of the annihilation chamber ($R =$~12~cm). Only events in which the total required variation, measured with a $\chi^2$-like variable~(with two degrees of freedom), did not exceed $\chi^2=0.5$ were retained for evaluation of the
$O_{\rm CPT}$ operator. This selection rejected poorly reconstructed events in which the geometrical uncertainty
of the estimated spin axis could deteriorate the resulting statistical polarisation of the o-Ps sample,
as well as spurious $3\gamma$ annihilation points resulting from reconstruction applied to background events (a comprehensive discussion of background sources is enclosed {in ``Methods''}).

For approximately two million selected events, the estimated spin axis direction and photon momentum vectors
were used to evaluate the $O_{\rm CPT}$ correlation~(equation~\ref{eq:operator}),
resulting in the distribution presented in Fig.~\ref{fig:panel}d.
Extraction of the degree of potential symmetry
violation from the angular correlation requires determination of the expectation value of the corresponding operator,
corrected by the analysing power of the setup. The latter is dominated by the effective ortho-positronium polarisation resulting from the evaluation of the spin axis orientation for each event. The effective polarisation is determined by
(i)~average longitudinal polarisation of positrons from \textsuperscript{22}Na~($P_{\text{e}^+}=\bar{\beta}\approx 0.67$),
(ii)~the amount of polarisation statistically transferred to the formed positronium~(2/3),
(iii)~ polarisation loss during positron thermalisation amounting to approximately~8~\%~\cite{cpt_positronium, yang}, and
(iv)~the geometrical uncertainty of the estimated direction of positron emission, which was reduced to approximately 9~\% in the presented setup~\cite{gajos_gps}. The total effective polarisation was estimated to be $P\approx 37.4$\~\%.

It should be noted that while previous experiments~\cite{Arbic:1988pv,Skalsey:1991vt,cp_positronium,cpt_positronium} concentrated on measuring the asymmetry of a given angular correlation, defined as $A(|O_{\rm CPT}|)=(N(+|O_{\rm CPT}|)-N(-|O_{\rm CPT}|))/(N(+|O_{\rm CPT}|)+N(-|O_{\rm CPT}|))$
~(where $N$ denotes the number of observed events for a given correlation value),
it is only a special case of the expectation value of the corresponding operator
in the case where a single value of $\pm|O_{\rm CPT}|$ is measurable simultaneously.

Therefore, we use the expectation value $<O_{\rm CPT}>$ over the entire region of its definition
as a measure of the observed asymmetry and obtain a value of
\begin{equation}
  \label{eq:asymmetry}
  <O_{\rm CPT}> = 0.00025 \pm 0.00036,
\end{equation}
thus observing no significant asymmetry.
The error is dominated by the statistical uncertainty of 0.00033.

For comparison with the measurements conducted to date, a parameter quantifying the level of observed CPT violation $C_{\rm CPT} \in [0,1]$ can be used, as in Ref.~\cite{cpt_positronium}, for which $C_{\rm CPT}=1$ corresponds to a maximal asymmetry violating CPT. Correcting the above result for the analysing power~($P$), we obtain

\begin{equation}
  \label{eq:ccpt}
  C_{\rm CPT} = <O_{\rm CPT}> / P = 0.00067 \pm 0.00095.
\end{equation}

\section*{{Discussion}}
We have demonstrated the application of tomographic reconstruction
of three-photon positronium annihilations
to determine the spin polarisation of ortho-positronium atoms,
enabling precise studies of discrete symmetries in positronium decays
without requiring a magnetic field.
Through the exploratory measurement presented in this work, we have demonstrated that the \jpet{} detector based on plastic scintillator strips can exclusively register three-photon annihilations of ortho-positronium atoms occurring in an extensive-size medium with reconstruction of the spatial location of the annihilation and o-Ps spin-axis estimation on a single-event basis. With this 26-day measurement, the sensitivity to the CPT-violating angular correlation between ortho-positronium spin and orientation of the annihilation plane
reached a statistical precision of $10^{-4}$, over a factor of three better 
than the previous result~\cite{cpt_positronium}.

Further sensitivity improvements to the presented CPT symmetry test are possible owing to an additional densely packed layer of plastic scintillators with a fully digital readout, which is currently being added to the \jpet{} detection setup.
Providing up to quadruple enhancement in registration probability for a single annihilation photon, it can increase the overall detection efficiency of \opsanh{} events by a factor of 64. This will be further augmented by a new design of the annihilation chamber with spherical geometry, increasing the o-Ps formation probability by a factor of approximately 1.5 with a given $\upbeta^+$~source activity and minimising potential spurious asymmetries from the geometry of the experimental setup~\cite{Gajos:2020symmetry}.
These improvements will allow the \jpet{} setup to collect
{the largest dataset of 
exclusively recorded ortho-positronium three-photon annihilations.}
Even with a moderate extension in measurement time compared to the experiment presented in this work, the statistics of the collected sample of \opsanh{} will allow for CPT symmetry testing at a precision level of $10^{-5}$.
Moreover, the control over systematic effects in the presented measurement
and coverage of a broad span of kinematic configurations of recorded events. This eliminates false asymmetries from the detector geometry, and opens the way to exploring
angular correlation operators for which usage was not feasible in previous experiments,
such as the $\mathbf{S}\cdot\mathbf{k}_{1}$ correlation, which is odd under both CP and CPT transformations~\cite{Gajos:2020symmetry}.

\section*{{Methods}}\label{sec:methods}

\subsection*{J-PET detector}\label{sec:meth:jpet}
The (\jpet{}) detector comprises three sparse concentric layers of 50~cm long plastic scintillator strips arranged axially in a cylindrical geometry~(see Fig.~\ref{fig:panel}a).
Each scintillator strip constitutes a detection module with two vacuum tube photomultipliers attached at its ends, which collect light produced as a result of Compton scattering of a $\gamma$~quantum in the strip.

The longitudinal positions and time of $\gamma$ interactions in the scintillator strips are reconstructed using times of light propagation from the point of $\gamma$ Compton scattering to the ends of the strip, providing interaction time resolution at a level of approximately 250~ps owing to the fast signals of the plastic scintillators.

Synchronisation of time signals from both sides of each scintillator, as well as between detection modules and layers, is obtained from self-calibration of the recorded data using a statistical technique exploiting time differences between $2\gamma$ annihilations and uncorrelated prompt photon emissions from the \textsuperscript{22}Na source~\cite{JPET_Calibration}.

\subsection*{Properties of the ortho-positronium production setup}\label{sec:ortho-positr-prod}

Density of porous material used in the experiment is estimated from densities
of pure silica gel (commonly accepted value
$d_{\mathrm{silica}}$
= 2.2~g~cm\textsuperscript{-3})
and gypsum,
$d_{\mathrm{gypsum}}$
= 2.3~g~cm\textsuperscript{-3}
. Density of mixture of both these components, taking into account their concentration, is slightly higher than that of pure silica gel and equal to ca.\ 2.22~g~cm\textsuperscript{-3}.
However, the so-called bulk density of coating composed of material particles, which we determined experimentally, is smaller and equal to 0.32~g~cm\textsuperscript{-3}.
This difference in the substrate density and the bulk density is related to the overall porosity of material and free interparticle spaces.

Textural characterization of material used in experiment was performed in a standard method on the basis of adsorption-desorption data of nitrogen at -196$^o$C.
Adsorption-desorption isotherms of nitrogen were measured using volumetric pore analyser ASAP 2040, Micromeritics Co.
Specific surface area SBET was derived from adsorption data in the range of relative pressure $p/p_{0}$~=~0--0.25 using Brunauer-Emmett-Teller equation~\cite{brunauer}.
The calculations were conducted under the assumption that one molecule of nitrogen occupies 0.162~nm$^2$.
Total pore volume Vp was estimated from a single point adsorption at $p/p_{0}=0.993$. Pore width DBJH and pore size distribution PSD were calculated using Barett-Joyner-Halenda procedure~\cite{barrett} applied to the desorption data. Numerical values of parameters characterizing investigated material are as follows: SBET = 252~m$^2$~g$^{-1}$, $V_{p}=0.58$~cm$^3$~g$^{-1}$ and average pore width DBJH = 6.6~nm. Relatively high specific surface area and total pore volume testifies that sample possess well developed system of pores and pore interior is accessible for nitrogen as an adsorptive. Hence, one can assume that pores are open to vacuum.

\subsection*{Identification of \opsanh{} annihilation event candidates}\label{sec:meth:identification}
\begin{figure}[h!]
  \centering
  \includegraphics[width=0.9\textwidth]{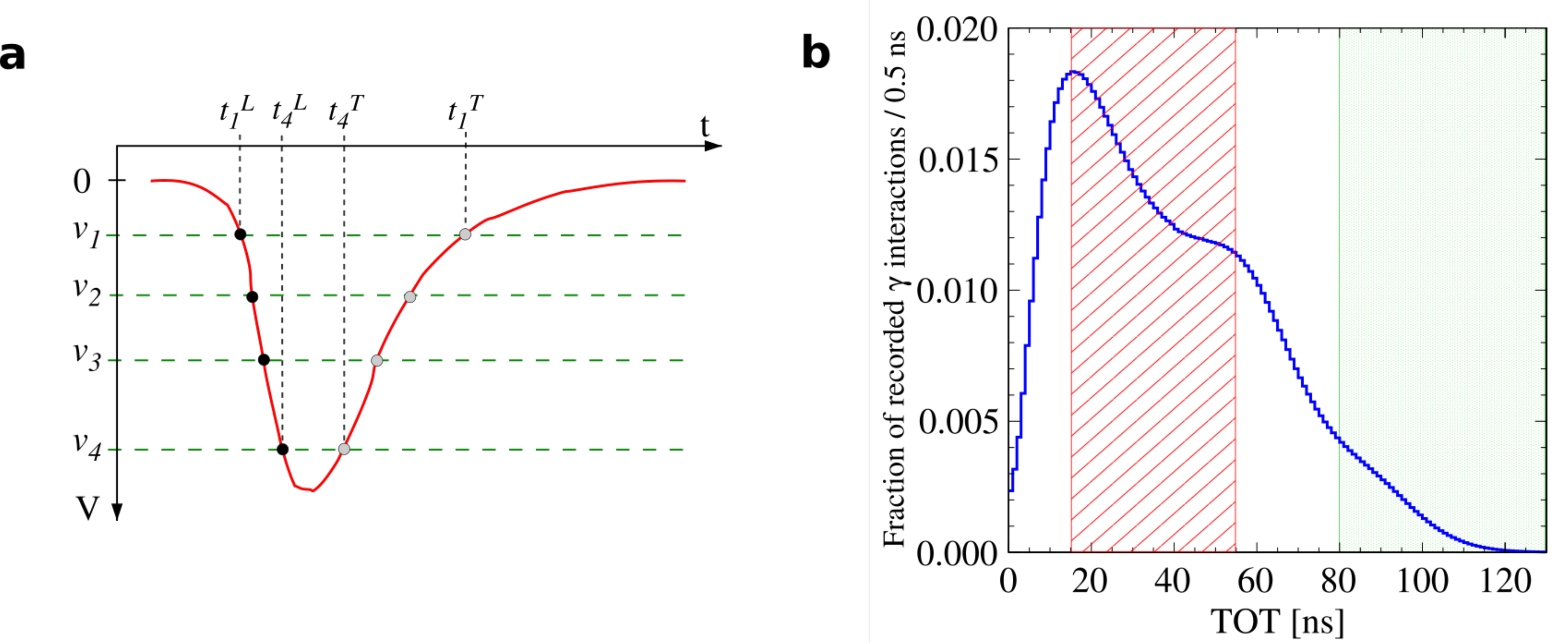}
  \caption{{\textbf{Measurement of deposited $\gamma$ energy with time-over-thresholds~(TOT).}}
    \textbf{a} Photomultiplier electric signals in \jpet{} sampled at four predefined voltage thresholds~{($\nu_1,\ldots,\nu_4$)},
    yielding four timestamps for the leading~($t_i^{\rm L}$) and trailing~($t_i^{\rm T}$) edges of the signal.
    The total TOT of a signal is calculated as $\sum_{i=1}^{4}(t_i^{\rm T}-t_i^{\rm L})$.
    \textbf{b} Distribution of total TOTs
    from both photomultiplier signals in \jpet{} detection modules, used as a measure of photon energy deposited in Compton scattering. The hatched red region is used to identify o-Ps annihilation photon candidates, whereas candidates for prompt photons from $^{22}$Ne* de-excitation are found in the green dotted region.}
  \label{fig:meth:tot}
\end{figure}

For every recorded interaction of a $\gamma$~quantum in the plastic scintillators,
the two photomultiplier electric signals are sampled by the \jpet{} front-end electronics in the voltage domain at four predefined amplitude thresholds applied to their leading and trailing edges~\cite{Palka:2017wms}, as depicted in Fig.~\ref{fig:meth:tot}a.
This technique provides a measure of the total size of the two signals through their TOTs, which is used to estimate 
the total energy deposited in the Compton scattering of the incident photon~\cite{Sharma:2020}.
Figure~\ref{fig:meth:tot}b presents the spectrum of TOT values for all recorded photon interactions,
in which the three visible structures correspond to an overlap of ortho-positronium annihilation and secondary Compton-scattered photons~(hatched red), Compton spectrum edge for 511~keV photons from e$^+$e$^-$~annihilations, and Compton edge for 1275~keV photons from de-excitation of the $\upbeta^+$ source~(dotted green), respectively, for increasing TOT values.
Candidates for ortho-positronium annihilation photons are selected by TOT values in a window of (15;55)~ns
to exclude noise and minimise the contribution of direct 2$\gamma$ annihilations that are abundant close to the middle Compton edge.

\begin{figure}[h!]
  \centering
  \includegraphics[width=0.9\textwidth]{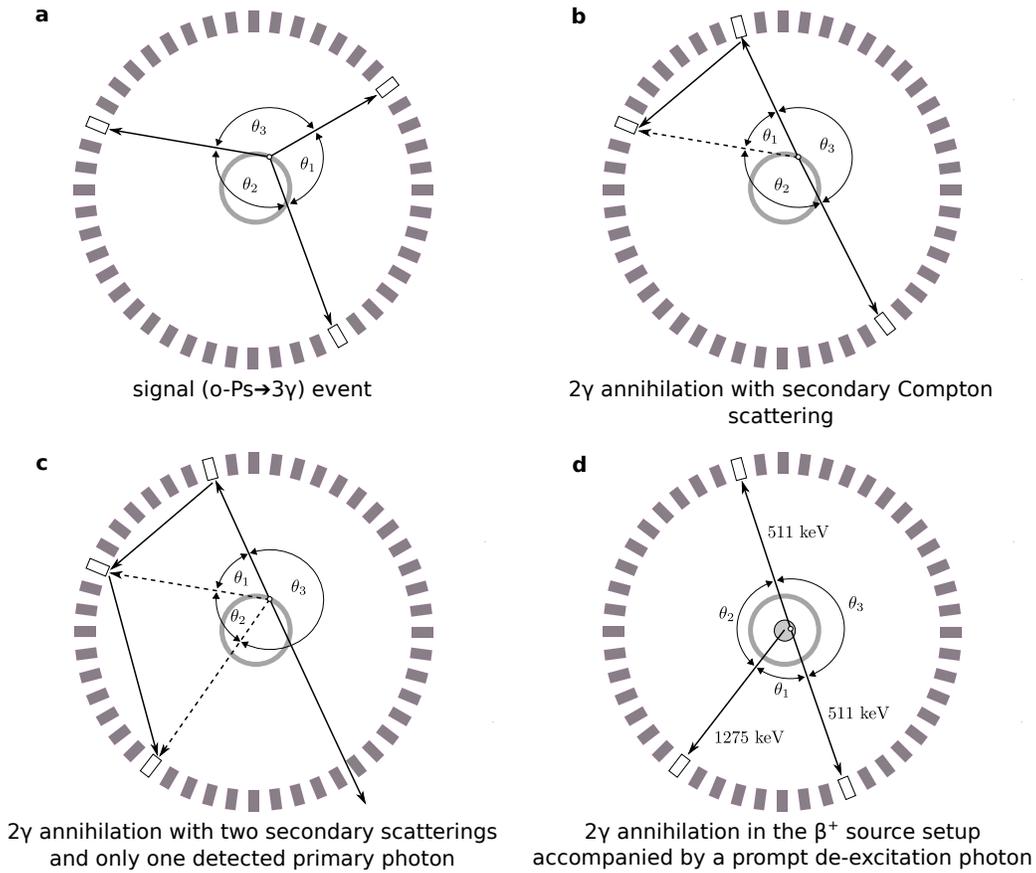}
  \caption{{\textbf{Schematic presentation of signal (o-Ps$\to 3\gamma$) and background events in the transverse view of the \jpet{} detector.}}
    The grey rectangles represent photon detection modules of plastic scintillators (only one layer is shown for readability). The modules in which photons were recorded are marked white.
      The annihilations take place in the wall of the vacuum chamber (grey band).
      In a signal event~\textbf{a}, all three photons from an ortho-positronium annihilation are recorded and the angles between their momenta, labelled according to ascending magnitude ($\theta_1 < \theta_2 < \theta_3$), obey $\theta_1+\theta_2 > 180^{\circ}$.
      The background is dominated by direct e$^+$e$^-\to 2\gamma$ annihilations, where the third recorded photon comes from secondary Compton scattering in the detector~\textbf{b}, resulting in the identification of a spurious primary photon marked with a dashed arrow.
      Three-photon events may also be caused by two subsequent secondary scatterings in the case where one of the direct annihilation photons is not detected~\textbf{c}, and by two-photon annihilations originating in the $\upbeta^+$~source setup~(grey circle) accompanied by a prompt \textsuperscript{22}Ne$^*$ de-excitation photon depositing low energy in Compton scattering~\textbf{d}.
    }
  \label{fig:event_schemes}
\end{figure}

Candidate events for o-Ps three-photon annihilations are identified as clusters of three-photon interactions recorded
in a coincidence time window of 2.5~ns.
At this stage, the background is mostly comprised of secondary Compton scatterings recorded in the detector, 
as well as 2$\gamma$~annihilations originating close to the $\upbeta^+$~source recorded alongside a low-energy deposition
of a de-excitation photon from the source itself.
The event topologies of the signal and major background components are presented schematically in Fig.~\ref{fig:event_schemes}. The next section describes the treatment of particular background sources.

\subsection*{Rejection of secondary Compton scatterings}\label{sec:meth:scattertest}
The use of plastic scintillators in J-PET entails the possibility of recording secondary Compton scatterings mimicking the registration of primary photons~(Figs.~\ref{fig:event_schemes}b,c).
Back-to-back e$^+$e$^-\to 2\gamma$ annihilations recorded with secondary scattering constitute a significant source of background for the identification of $3\gamma$ events.
However, the time resolution of plastic scintillators also allows such events to be distinguished by testing a hypothesis of a photon propagating between every two recorded interactions, quantified by
\begin{equation}
  \label{eq:scat-test}
  \delta_{ij} = |d_{ij} - c\Delta t_{ij}| = \left| |\mathbf{r}_i-\mathbf{r}_j| - c\left|t_i-t_j\right| \right|,
\end{equation}
where $d_{ij}$ is the distance between the $i$th and $j$th photon interactions recorded in an event candidate 
with recording times of $t_i$ and $t_j$, respectively,
and $c$ denotes the velocity of light.

\begin{figure}[h!]
  \centerline{
      \includegraphics[width=0.5\textwidth]{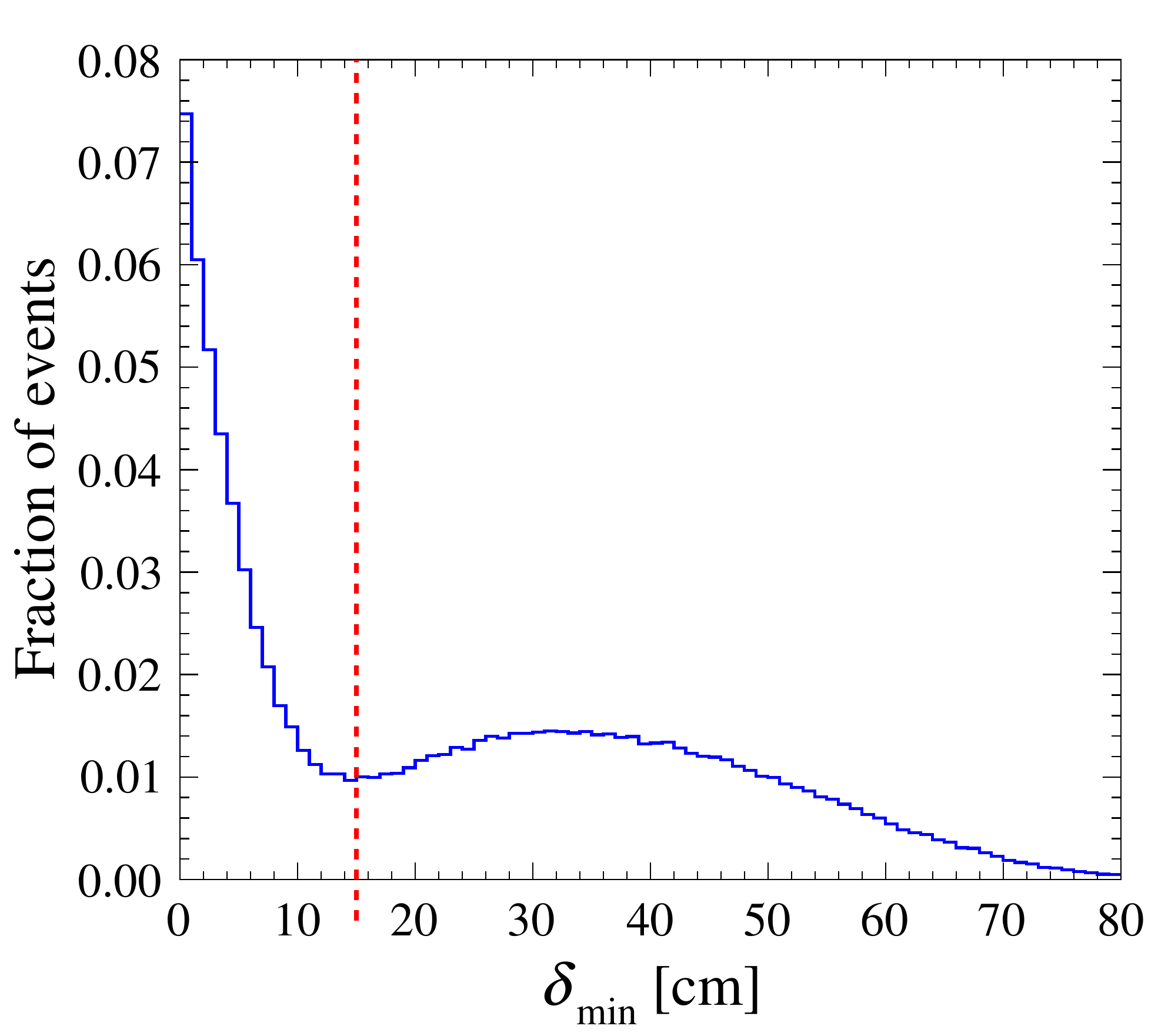}      
  }
  \caption{{\textbf{Rejection of secondary Compton scatterings.}}
    Distribution of the minimal discrepancy between the travelled path and time of flight among all hypothetical secondary scattered photons considered in a single $3\gamma$~event candidate. Events with $\delta_{\rm min}{<}15$~cm~(marked with the red line) are discarded as containing secondary Compton scatterings in the detector modules.
  }
  \label{fig:meth:scatter-test}
\end{figure}

With such definition, $\delta_{ij}$ assumes values close to zero for photon interaction pairs
corresponding to a secondary Compton scattering.
Consequently, its contamination among three-photon events is identified using
$\delta_{\rm min}=\min\{\delta_{ij}\}$ for $i{\neq}j{=}1,2,3$ calculated for every event.
The contribution of events with secondary scattering, clearly visible in Fig.~\ref{fig:meth:scatter-test}, is rejected as events with $\delta_{ij} < 15$~cm are discarded.

\begin{figure}[h!]
  \centerline{
    \includegraphics[width=0.9\textwidth]{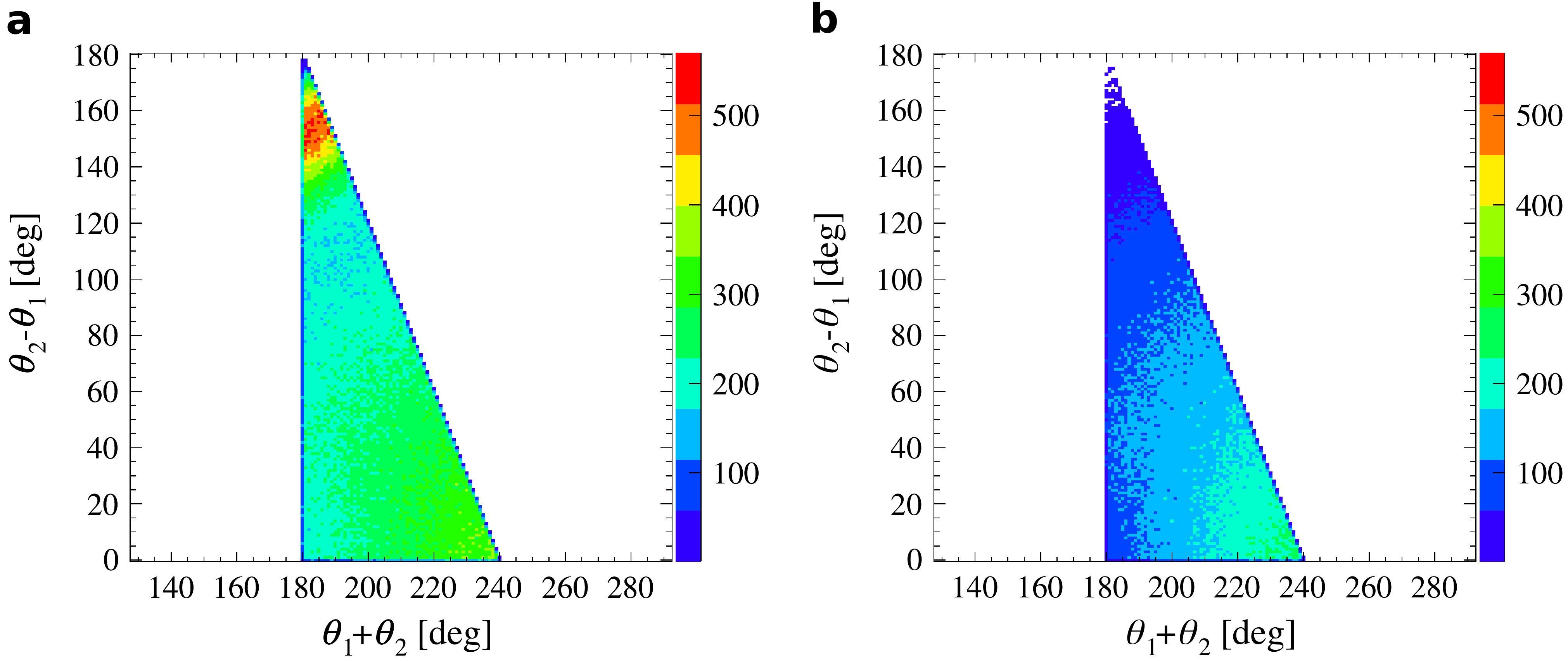}      
  }

  \caption{{\textbf{Effect of secondary Compton scatterings' rejection.}}
    Relative distribution of the sum and difference of the two smallest angles between photon momenta
    (see Fig.~\ref{fig:event_schemes})
    in the identified $3\gamma$~events before (\textbf{a}) and after (\textbf{b}) rejection of the secondary scattered photons.
    The $\theta_1+\theta_2>180^{\circ}$ limit results from momentum conservation in \opsanh{} annihilations.
    o-Ps annihilation events are expected in the right corner of the populated region~\cite{Kaminska:2015yqa},
    whereas the topmost region is specifically for secondary scattered photon events.  
}
  \label{fig:meth:angles}
\end{figure}

The effect of the removal of secondary Compton scatterings is visible in the relative distribution of the sum and difference of the two smallest angles between the reconstructed momenta of the three photons, which is a convenient figure for evaluation
of the \opsanh{} sample purity~\cite{Kaminska:2015yqa}.
As follows from a comparison of schemes a and b--d in Fig.~\ref{fig:event_schemes}, the genuine 3$\gamma$ annihilations are characterised by large sums of the two smallest angles $\theta_1+\theta_2$ with a relatively small difference between them. Conversely, secondary scattering events are located closer to $\theta_1+\theta_2\approx 180^{\circ}$ with a large $\theta_2-\theta_1$ value. Figure~\ref{fig:meth:angles} presents the relative distribution of these figures 
before and after application of the criterion on $\delta_{\rm min}$, demonstrating the purification of the event sample
from secondary scattering events.

\subsection*{Rejection of two-photon annihilations coincident with a de-excitation photon}\label{sec:dmin}
As indicated in Fig.~\ref{fig:panel}b, the positron source used in this experiment emits prompt photons
of 1275~keV owing to de-excitation of the $\upbeta^+$ decay product. While such photons provide a useful start signal
for evaluation of the positronium lifetime,
their Compton interactions deposit energy from a continuous spectrum, which may mimic low-energy annihilation photons.
As a result, the identified \opsanh{} event candidates are contaminated by two-photon direct e$^+$e$^-$~annihilations accompanied by a de-excitation $\gamma$~quantum if it deposits low energy in a \jpet{} detection module~(Fig.~\ref{fig:event_schemes}d).
This background is recognised by the difference in azimuthal coordinates of the detection modules~(labelled $\theta^{\rm XY}$) between two of the three interaction points being close to $180^{\circ}$.
Moreover, if hypothetical lines of response are considered for every pair of recorded $\gamma$ interactions
and tomographic reconstruction of a $2\gamma$ annihilation point is performed, such events are characterised
by one of the pairs yielding a hypothetical annihilation point at a small distance $d_{\rm LOR}$ from the $\upbeta^+$~source.

\begin{figure}[h!]
  \centering
  \includegraphics[width=0.5\textwidth]{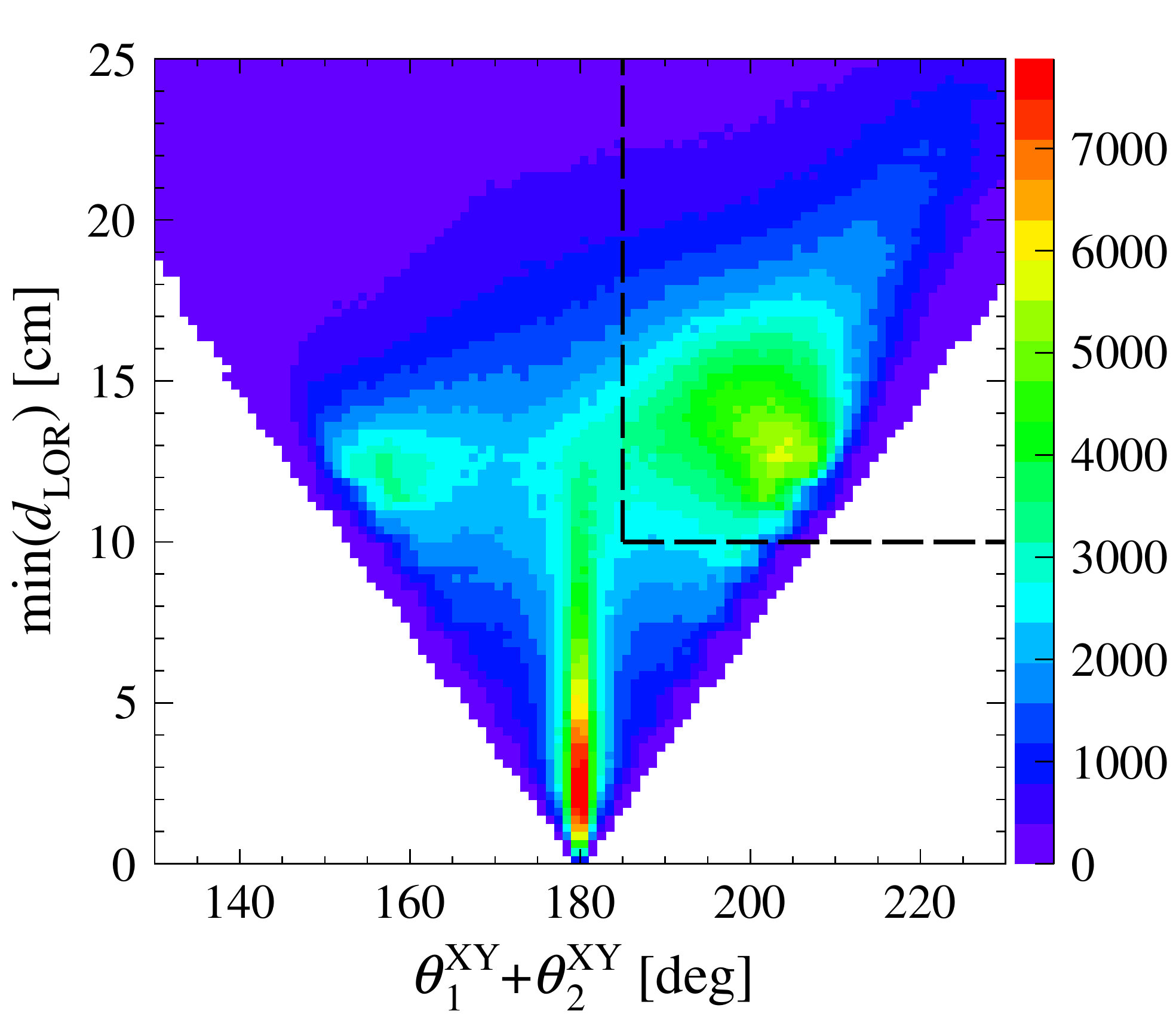}
  \caption{{\textbf{Separation of ortho-positronium annihilations from direct two-photon events.}}
    Distribution of distance between the $\upbeta^+$ source location and the closest hypothetical 2$\gamma$ annihilation point on a line of response between two recorded photon interactions vs. the sum of the two smallest angles between azimuthal coordinates of the recorded $\gamma$ interaction points.
    Events located in the signal-specific upper right region of the distribution (marked with dashed black line) are retained for the analysis in order to discriminate background events from 2$\gamma$ annihilations (see Fig.~\ref{fig:event_schemes}b--d).
    }
  \label{fig:dmin_angles}
\end{figure}

Figure~\ref{fig:dmin_angles} presents the minimal values of $d_{\rm LOR}$ in each three-photon event as a function of the sum of the two smallest azimuthal locations of recorded $\gamma$ interactions, where a structure corresponding to background events from direct e$^+$e$^-$~annihilations accompanied by a prompt photon originating in the source setup are clustered around $180^{\circ}$ and small values of $d_{\rm LOR}$.

A small contribution of direct 2$\gamma$ annihilations accompanied by a scattered photon which survive {the previous selection steps} is also revealed in this distribution at about $\theta_1^{\rm XY}+\theta_2^{\rm XY}\approx160^{\circ}$. Since genuine \opsanh{} events are mostly characterized by $\theta_1^{\rm XY}+\theta_2^{\rm XY} > 180^{\circ}$ and large $d_{\rm LOR}$,  events located in the region marked with dashed black line in Fig.~\ref{fig:dmin_angles} are selected for the final analysis.

\subsection*{Evaluation of systematic uncertainties}\label{sec:systematics}
False asymmetries can originate from the geometry of the positronium production setup
or the detector,
as well as from the contributions of background processes such as cosmic radiation recorded during measurement.
Geometrical effects were evaluated by determining the following CPT-even and CP-odd angular correlation operator:
\begin{equation}
  O_{\rm CP} = (\mathbf{S}\cdot\mathbf{k}_1)(\mathbf{S}\cdot\mathbf{k}_1\times\mathbf{k}_2).  
\end{equation}
While it is potentially sensitive to CP-violation effects~\cite{Bernreuther:1988tt,cp_positronium},
these could only be observed in the case of a tensor-polarised ortho-positronium sample available when an external magnetic field~($\mathbf{B}$) was used.
As no $\mathbf{B}$ fields were present in the \jpet{} measurements, all asymmetries manifested in the distribution
of $O_{\rm CP}$ must have originated from the detector geometry.
The mean value of this operator was found to be
(1.3$\pm1.4_{\rm stat})\times 10^{-4}$,
showing no significant detector asymmetries at the level of $10^{-4}$.

The impact of cosmic rays recorded during the experiment was estimated in a dedicated measurement without the $\upbeta^+$ source and positronium annihilation setup, where only cosmic radiation was recorded.
It was estimated that cosmic ray events passing the event selection criteria may constitute at most 3.9$\times 10^{-5}$ of the final event sample, which was considered as part of the systematic uncertainty.

The influences of subsequent event selection criteria applied in the analysis were tested by variation of each of the cut values within a few multiples of their corresponding experimental resolution.
The only analysis cut found to have a statistically significant impact on the measured
value of $\langle O_{\rm CPT}\rangle$
was the upper bound on the TOT of candidates for annihilation photon interactions
(see Fig.~\ref{fig:meth:tot}b).
Consequently, an effect of $1.26\times 10^{-4}$ on $\langle O_{\rm CPT}\rangle$, corresponding to a variation
of this bound by 1~ns, was included in the systematic uncertainty.


\section*{Acknowledgements}
\label{ack}
The J-PET collaboration acknowledges the support provided by the Polish National Center for Research and Development through grant
INNOTECH-K1/IN1/64/159174/NCBR/12;
the Foundation for Polish Science through the MPD and
TEAM POIR.04.04.00-00-4204/17 
Programmes; the National Science Centre of Poland through grant nos.
2016/21/B/ST2/01222,
2017/25/N/NZ1/00861, and
2019/35/B/ST2/03562;
the Ministry for Science and Higher Education through grant nos.
6673/IA/SP/2016 and
7150/E- 338/SPUB/2017/1;
the Jagiellonian University via project no. CRP/0641.221.2020,
the Austrian Science Fund FWF-P26783.
as well as the SciMat and DigiWorld Priority Research Area budgets under the programme Excellence Initiative - Research University at the Jagiellonian University.

\section*{Author contributions}
The experiment was conducted using the J-PET apparatus developed by the J-PET collaboration.
The J-PET detector and the techniques of the experiment were conceived by P.M.
The data analysis was conducted by A.G. and M.M.
The positronium production chamber was designed and constructed by M.G., J.G., B.J., and A.S.
{
  P.~M.,
  A.~G.,
  M.~M.,
  J.~C.,
  N.~C.,
  C.~C.,
  E.~C.,
  M.~D.,
  K.~D.,
  B.~C.~H.,
  \L{}.~K.,
  H.~K.,
  D.~K.,
  G.~K.,
  N.~K.,
  T.~K.,
  E.~K.,
  S.~N.,
  S.~P.,
  M.~P.~N.,
  J.~R.,
  S.~S.,
  Shivani,
  M.~Silarski,
  M.~Skurzok,
  E.~\L{}.~S.
  and
  F.~T.
  participated in the construction, commissioning and operation of the experimental setup.
  K.~D.,
  A.~G.,
  K.~Kacprzak,
  K.~Klimaszewski,
  P.~K.,
  N.~K.,
  W.~K.,
  S.~N.,
  L.~R.,
  R.~Y.~S.,
  and
  W.~W.
  participated in the development of data reconstruction and analysis procedures.
}
  The manuscript was prepared by P.M and A.G and was then edited and approved by all authors.

\section*{Data availability}
{The datasets collected in the experiment and analysed during the current study are available
under restricted access due to the large data volume.
Direct access to the data can be arranged on a reasonable request
by contacting the corresponding authors.
}

\section*{Competing interests}
The authors declare no competing interests.



\end{document}